\documentclass[11pt]{article}

\usepackage[utf8]{inputenc}
\usepackage{amsmath}
\usepackage{amssymb}
\usepackage{amsthm}
\usepackage{xspace}
\usepackage{xcolor}
\usepackage{graphicx}
\usepackage{booktabs}
\usepackage{multirow}
\usepackage{tabularx}
\usepackage{mdframed}
\usepackage{listings}
\usepackage[ruled,vlined]{algorithm2e}
\usepackage{svg}
\usepackage{float}
\usepackage{subcaption}
\usepackage{enumitem}

\definecolor{light-gray}{gray}{0.95}
\definecolor{conceptblue}{rgb}{0.2,0.2,0.7}
\definecolor{negationred}{rgb}{0.6,0.1,0.1}
\definecolor{logicgreen}{rgb}{0.1,0.5,0.1}

\usepackage[colorlinks=true,
            linkcolor=blue,
            citecolor=blue,
            urlcolor=blue]{hyperref}

\lstdefinestyle{fuzzyquery}{
    basicstyle=\ttfamily\footnotesize,
    backgroundcolor=\color{white},
    keywordstyle=\color{logicgreen}\bfseries,
    commentstyle=\color{gray},
    stringstyle=\color{conceptblue},
    morekeywords={AND, OR, NOT, DEFINE, CLASS, AS},
    moredelim=**[is][\color{conceptblue}]{@@}{@@},
    breaklines=true,
    frame=single,
    tabsize=4,
    showstringspaces=false
}

\SetKwInput{KwInput}{Input}
\SetKwInput{KwOutput}{Output}

\title{Fuzzy Ontology Embeddings and Visual Query Building for Ontology Exploration}

\author{
\begin{tabular}{cc}
Vladimir Zhurov\thanks{\texttt{vladimir.zhurov@uwo.ca}} & John Kausch\thanks{\texttt{jkausch@uwo.ca}} \\
\small Department of Computer Science & \small Faculty of Information and Media Studies\\
\small University of Western Ontario & \small University of Western Ontario\\[0.5em]
Kamran Sedig\thanks{\texttt{kamrans@uwo.ca}} & Mostafa Milani\thanks{\texttt{mostafa.milani@uwo.ca}} \\
\small Department of Computer Science \& & \small Department of Computer Science \\
\small Faculty of Information and Media Studies & \small University of Western Ontario\\
\small University of Western Ontario &
\end{tabular}
}

\date{} 

\begin{document}

\newcommand{\red}[1]{\textcolor{red}{#1}}
\newcommand{\blue}[1]{\textcolor{blue}{#1}}
\newcommand{\purple}[1]{\textcolor{purple}{#1}}

\newcommand{\mostafa}[1]{\red{Mostafa:#1}}
\newcommand{\vlad}[1]{\blue{Vlad:#1}}
\newcommand{\todo}[1]{\purple{TODO:#1}}

\newcommand{\ignore}[1]{}

\newtheorem{example}{Example}

\newcommand{\cneed}[1]{\red{[CITATION NEEDED:#1]}}
\newcommand{\nit}[1]{\textsf{#1}}

\newcommand{\boxtheorem}{\hfill $\blacksquare$\vspace{2mm}}

\newcommand{\bigcomment}[1]{}

\newcommand{\alc}{$\mathcal{ALC}$\xspace}
\newcommand{\falcon}{\textsc{FALCON}\xspace}
\newcommand{\query}{\mathcal{Q}\xspace}
\newcommand{\Q}{\mathcal{Q}\xspace}
\renewcommand{\O}{\mathcal{O}\xspace}
\newcommand{\C}{\mathcal{C}\xspace}
\newcommand{\U}{\mathcal{U}\xspace}
\newcommand{\mc}[1]{\mathcal{#1}\xspace}
\newcommand{\similar}{s}

\renewcommand{\labelitemi}{$\bullet$}   
\renewcommand{\labelitemi}{\scalebox{0.5}{$\blacksquare$}}   
\renewcommand{\labelitemii}{$\circ$}   
\renewcommand{\labelitemiii}{$\star$}  
\renewcommand{\labelitemiv}{$\diamond$} 

\newcommand{\roles}{\mathcal{R}}
\newcommand{\concepts}{\mathcal{C}}
\newcommand{\inds}{\mathcal{N}}
\newcommand{\abox}{\mc{A}}
\newcommand{\tbox}{\mc{T}}
\newcommand{\ontology}{\mc{O}}
\newcommand{\similarities}{\text{sims}}
\newcommand{\Embedding}{\text{Embedding}}
\newcommand{\floqe}{\text{FuzzyVis}\xspace}
\newcommand{\eFunc}{f}

\newcommand{\tnorm}{\theta}
\newcommand{\tconorm}{\kappa}
\newcommand{\fneg}{\eta}

\maketitle

\begin{abstract}
Ontologies play a central role in structuring knowledge across domains, supporting tasks such as reasoning, data integration, and semantic search. However, their large size and complexity—particularly in fields such as biomedicine, computational biology, law, and engineering—make them difficult for non-experts to navigate. Formal query languages such as SPARQL offer expressive access but require users to understand the ontology’s structure and syntax. In contrast, visual exploration tools and basic keyword-based search interfaces are easier to use but often lack flexibility and expressiveness.

We introduce {\em \floqe}, a proof-of-concept system that enables intuitive and expressive exploration of complex ontologies. \floqe\ integrates two key components: a fuzzy logic-based querying model built on fuzzy ontology embeddings, and an interactive visual interface for building and interpreting queries. Users can construct new {\em composite concepts} by selecting and combining existing ontology concepts using logical operators such as conjunction, disjunction, and negation. These composite concepts are matched against the ontology using fuzzy membership-based embeddings, which capture degrees of membership and support approximate, concept-level similarity search.

The visual interface supports browsing, query composition, and partial search without requiring formal syntax. By combining fuzzy semantics with embedding-based reasoning, \floqe\ enables flexible interpretation, efficient computation, and exploratory learning. A usage scenario demonstrates how \floqe\ supports subtle information needs and helps users uncover relevant concepts in large, complex ontologies.
\end{abstract}


\section{Introduction}\label{sec:intro}

Ontologies are formal representations of knowledge that describe the concepts in a knowledge domain, their relationships, and associated instances. They provide a structured framework for representing and reasoning about knowledge, supporting areas such as knowledge representation, artificial intelligence (AI), and data management. In knowledge representation and AI, ontologies help standardize terminology and organize domain knowledge in large datasets, enabling automated reasoning tasks such as inference, problem-solving, and theorem proving~\cite{arbabi2019identifying,eiter2006reasoning,hayes2013second}. They often serve as advanced thesauri for query expansion, improving search effectiveness and interoperability~\cite{azad2019query, asfand2020semantic}. In data management, ontologies are central to data integration by supporting the unification of heterogeneous data sources~\cite{de2018using} and enhancing concept recognition. A key approach is {\em ontology-based data access}, which uses ontologies to enable reasoning over large datasets---typically stored in databases---to answer complex queries~\cite{calvanese2011mastro}.

Ontologies are widely used to represent structured knowledge across various domains. In biology, the {\em Gene Ontology (GO)}~\cite{gene2000go} and the {\em Sequence Ontology (SO)}~\cite{eilbeck2005sequence} provide standardized annotations for genes and sequences. In medicine, the {\em Human Phenotype Ontology (HPO)}~\cite{human2024nucl}, the {\em Disease Ontology (DO)}~\cite{schriml2019disease}, and {\em MeSH (Medical Subject Headings)}~\cite{lipscomb2000medical} support standardized indexing and search of biomedical literature and clinical data. In the legal domain, ontologies such as {\em LKIF-Core}~\cite{broekstra2009lkif}, {\em LODE}~\cite{boer2009lode}, and {\em ECLAP}~\cite{rigas2012eclap} support legal reasoning and document management. When no ontology exists for a given domain, one can often be developed to capture its core concepts and relationships.

A persistent challenge in working with ontologies is their size and complexity. Many widely used ontologies---such as GO and HPO---contain hundreds of thousands of concepts, making them difficult to navigate, search, and understand, particularly for non-expert users. Using such ontologies effectively requires {\em exploratory learning}, a process through which users build mental models by identifying key concepts, understanding their relationships, and uncovering structural patterns~\cite{demelo2021forming}. This process often involves locating {\em landmarks} (central or well-known concepts), following {\em routes} (chains of related concepts), and exploring {\em neighborhoods} (clusters of semantically related concepts).

Exploratory learning over ontologies often involves two core functionalities: searching for relevant concepts and querying to express complex information needs. Search is typically based on keyword matching against concept labels or textual descriptions in the ontology. While simple to use, such methods frequently fail due to vocabulary mismatch, ambiguity, or variation in phrasing. On the other end of the spectrum, formal query languages such as SPARQL~\cite{Seaborne13SQL}, OWL-QL~\cite{fikes2004owlql}, SeRQL~\cite{broekstra2002serql}, and SPARQL-DL~\cite{sirin2007sparql} offer expressive power but require users to understand both the ontology’s structure and the syntax of the query language. These limitations make them difficult to use, especially for non-experts, and the results often require further interpretation. As a result, users exploring large ontologies are left with few practical options when their information needs are vague, evolving, or difficult to articulate precisely.

To address this gap, we propose a querying approach that strikes a balance between expressiveness and simplicity. Instead of writing formal queries, users define new concepts by composing existing primitive concepts in the ontology using set operators---conjunction, disjunction, and negation. These user-defined composite concepts are treated as fuzzy queries and are compared against the ontology to retrieve the most semantically similar primitive concepts. This supports intuitive interaction, flexible query construction, and approximate matching---even when the target composite concept is not explicitly defined in the ontology or when the query is underspecified or partially conflicting. Our approach enables users, particularly non-experts, to express their intent using simple, interpretable building blocks while still benefiting from the structure and semantics of the ontology. We now illustrate this with an example.

\begin{example} \label{ex:intro}\em To illustrate our querying approach, consider a physician, whom we refer to as the user, exploring HPO to investigate a patient's symptoms: difficulty speaking and swallowing, with no indication of immune dysfunction. The user suspects a neuromuscular issue but is unsure of the precise terminology.

Through browsing and personal knowledge, the user finds three relevant primitive concepts: \nit{Slurred speech} (\nit{HP\_0001350}), \nit{Dysphagia} (difficulty or discomfort in swallowing, \nit{HP\_0002015}), and \nit{Abnormality of the immune system} (\nit{HP\_0002715}). Using our interface, they define a composite query concept \( Q \):%
\begin{equation*}
Q \equiv \nit{Slurred speech} \sqcap \nit{Dysphagia} \sqcap \neg \nit{Abnormality of the immune system}.
\end{equation*}

\noindent This concept captures the co-occurrence of speech and swallowing difficulties while explicitly excluding immune-related causes. Such a combination is not explicitly defined in the ontology and would typically yield no results using standard (non-fuzzy) query tools.

Using our fuzzy querying approach, the system interprets this as a soft concept and returns the most similar known phenotypes. These may include \nit{Pseudobulbar paralysis} (\nit{HP\_0007024}), referring to a neurological condition, which exhibits some of the patient's symptoms, or \nit{Abnormal esophagus physiology} (\nit{HP\_0025270}), referring to structural abnormalities of the patient's esophagus, which could lead to food entering the air way. Both are relevant concepts to the user's intent and could be difficult to find through keyword search alone.\boxtheorem
\end{example}

To implement our querying approach, we represent ontologies as fuzzy ontologies with fuzzy interpretations, where each concept is treated as a fuzzy set assigning a membership degree to each element in the interpretation domain. Given an ontology, we use existing fuzzy reasoners to construct such interpretations that approximately satisfy the ontology's axioms. Based on these interpretations, we introduce a novel method for generating ontology embeddings, where each concept is represented as a vector of membership degrees over a fixed, ordered set of domain elements. This allows us to precompute embeddings for all primitive concepts offline. At query time, when a user defines a composite concept using conjunction, disjunction, or negation over known primitives, its embedding is computed by applying the corresponding fuzzy operations element-wise to the membership vectors. This compositional capability enables flexible query construction and is not supported by existing ontology embedding methods. The resulting embeddings also support efficient similarity computations (e.g., cosine similarity) between user-defined concepts and existing primitive concepts in the ontology, enabling approximate matching even when no exact concept exists.

We implement this querying approach in {\em \floqe}, a visual system for ontology exploration. In addition to standard features such as keyword search and a typical ontology views, \floqe\ provides a complex interactive visualization with features such as fisheye distortion and annotation for navigating large ontologies. Users can build visual queries by collecting relevant primitive concepts and combining them using logical operations such as conjunction, disjunction, and negation. The system computes the embedding of the resulting composite concept and returns the most similar primitive concepts as suggestions. Beyond querying, \floqe\ supports interactive exploration by linking search, query results, and navigation. Concepts returned in results can be located in the primary visualization, and collected concepts can be reused across queries. By combining fuzzy ontology embeddings, query answering, and an intuitive visual interface, \floqe\ enables non-experts to query and explore large ontologies effectively. It supports exploratory learning through flexible query construction, approximate matching, and structural similarity, addressing key limitations of existing ontology querying tools.

The paper is organized as follows. Section~\ref{sec:related} reviews related work. Section~\ref{sec:preliminaries} introduces the necessary preliminaries and background concepts. Section~\ref{sec:main} presents our fuzzy ontology embeddings and describes our querying approach based on these embeddings. Section~\ref{sec:floqe} details the \floqe\ system, its implementation, and practical use cases. Section~\ref{sec:conclusion} concludes the paper and outlines future directions.
\section{Related Work}\label{sec:related}

This section reviews related work on ontology embedding methods (Section~\ref{sec:rw-emb}) and existing tools for ontology visualization and exploration (Section~\ref{sec:rw-visual}).

\subsection{Ontology Embeddings} \label{sec:rw-emb}

To support scalable computation over ontologies, numerous methods embed ontology elements into continuous spaces that preserve semantic relationships and enable downstream tasks such as classification, clustering, reasoning, and search~\cite{chen2025ontology}. Ontology embedding refers to the process of mapping elements of an ontology to structured mathematical representations---typically vectors, regions, or lattices---that reflect their semantics. Formally, an embedding function \(\eFunc\) maps each ontology element \(e \in \ontology\) (e.g., a concept, relationship, or individual) to a vector \(\eFunc(e) = \mathbf{v}_e \in \mathbb{R}^d\) in a \(d\)-dimensional space. Below, we review the main families of embedding methods, organized by their modeling approach:%
\begin{itemize}
    \item {\em Translation-based embeddings.} These methods model relations as translation operations in vector space. Given a triple \((h, r, t)\)---where \(h\) and \(t\) are individuals (head and tail), and \(r\) is a relation---the goal is to learn embeddings such that \(\mathbf{v}_h + \mathbf{v}_r \approx \mathbf{v}_t\). This approach was introduced by TransE~\cite{bordes2013translating}, which assumes relational meaning can be captured by a fixed vector offset. Extensions such as TransH~\cite{wang2014knowledge} and TransR~\cite{lin2015learning} introduce projection mechanisms to better handle complex relations, embedding entities and relations in different spaces. While scalable and widely used for knowledge graph completion, these models are limited to triple-based structures and struggle with more complex ontology constructs such as concept hierarchies and logical combinations.
    
    \item {\em Path-based and lexical methods.} These methods extract sequential patterns from ontologies or knowledge graphs and apply language modeling techniques to learn embeddings. The key idea is that concepts or entities frequently co-occurring along paths in the graph are likely to be semantically related. RDF2Vec~\cite{ristoski2016rdf2vec} performs random walks over RDF graphs to generate sequences, which are embedded using word2vec. OWL2Vec~\cite{chen2021owl2vec} extends this to OWL ontologies by incorporating logical axioms, syntactic features, and lexical information across multiple ontology projections (e.g., logical, lexical, assertional). These methods are interpretable and scalable, especially for lightweight ontologies with rich textual annotations. However, they often lack alignment with formal semantics and cannot guarantee logical consistency.
    
    \item {\em Geometric and neurosymbolic embeddings.} These methods represent ontology elements as geometric objects (e.g., vectors, balls, boxes, or cones) in continuous space and use geometric operations to simulate logical reasoning~\cite{alivanistos2022neurosymbolic}. The intuition is that logical constructs such as subsumption, conjunction, and disjunction can be modeled using spatial relationships (e.g., containment or intersection). For example, EL Embeddings~\cite{kulmanov2019embeddings} represent concepts as Euclidean balls, where subsumption corresponds to containment. BoxE~\cite{jackermeier2024dual} uses axis-aligned boxes to capture more complex logical structures. Hyperbolic models such as ConeE~\cite{chen2024ontology} embed taxonomies as convex cones to preserve hierarchical structure. RotatE~\cite{sunrotate}, though originally designed for knowledge graphs, uses complex rotations to capture properties like symmetry and inversion. These models offer strong logical expressiveness but are often computationally intensive and less suited for dynamic or user-defined query concepts.

\end{itemize}

In addition to vector-based methods, some models use algebraic structures such as lattices or order embeddings to capture subsumption hierarchies and logical entailment~\cite{zhapa2023cate}. While expressive and theoretically grounded, these approaches are typically less practical. 

The existing ontology embedding methods we reviewed above are designed for static ontologies with fixed concept sets and offer limited or no support for composing concepts using logical operations. For example, geometric models such as BoxE~\cite{jackermeier2024dual} and ConeE~\cite{chen2024ontology} can approximate conjunction using the intersection of regions. However, they do not support negation and offer limited or no support for disjunction. Since these methods require retraining to incorporate new concepts, they are not suited for dynamic, query-time composition. This makes them incompatible with our approach, where users define new composite concepts at runtime and embeddings must be computed on the fly without retraining. Our embedding method addresses this limitation by enabling on-the-fly construction of concept embeddings through fuzzy operations, supporting efficient and flexible concept-level search over large-scale ontologies.

\subsection{Ontology Visualization and Exploration Tools} \label{sec:rw-visual}

Ontology visualization tools are designed to help users explore, understand, and query complex ontologies. These tools typically fall into three categories: 
\begin{itemize}
    \item {\em Indented tree views} display concept hierarchies in a collapsible format, making them intuitive for browsing taxonomies. Tools such as Kaon~\cite{gabel2004d3} support ontology editing and provide treemap-based navigation. WebProtégé~\cite{musen2015protege}, a widely used collaborative editor, also adopts a tree-based layout for concept exploration. OBO-Edit~\cite{day2007obo}, designed for biomedical ontologies, uses a similar structure. While effective for viewing hierarchical relationships, these tools are less suitable for exploring complex inter-concept connections or non-hierarchical structures.
    
    \item {\em Graph-based visualizations}, also known as node-link diagrams, represent ontology elements as nodes and their relationships as edges. This approach is better suited for capturing both hierarchical and non-hierarchical semantics. Tools such as OWLViz~\cite{lohmann2014protegevowl} and NavigOWL~\cite{hussain2014scalable} extend Protégé with interactive graph views. TGVizTab~\cite{alani2003tgviztab}, OWLPropViz~\cite{wachsmann2008owlpropviz}, and OntoGraf allow users to dynamically expand and collapse nodes. WebVOWL~\cite{lohmann2015webvowl} brings this functionality to the browser, providing interactive, user-friendly visualizations based on the VOWL specification. While graph-based tools reveal richer relational context, they often struggle with visual clutter as ontology size increases.
    
    \item To improve scalability and clarity, several tools combine tree and graph layouts or adopt alternative visual metaphors. Jambalaya~\cite{lintern2005jambalaya}, a Protégé plugin, integrates hierarchical and relational views with enhanced layout control. PRONTOVISE~\cite{demelo2021forming} supports multiple visual modes, including bar charts, to compare concept properties. CODEX~\cite{hartung2012codex} focuses on ontology evolution and uses pie charts and word clouds to illustrate changes across versions. OntoTrix~\cite{bach2011ontotrix} replaces node-link diagrams with adjacency matrices to reduce clutter in dense graphs. More recently, Graffinity~\cite{smith2020graffinity} has introduced a hybrid approach that combines matrix and node-link views to support the visualization of large ontologies.
\end{itemize}

In summary, tree-based tools are well-suited for browsing hierarchies, while graph-based and hybrid methods offer broader relational context. Although recent tools address scalability to some extent, visualizing and querying large ontologies in an interactive, user-friendly manner remains a challenge~\cite{dudavs2018ontology}.

Most existing tools are primarily designed for navigating fixed hierarchies or inspecting predefined relations. They offer limited support for composing flexible, user-defined queries that involve multiple semantic conditions, such as conjunction, exclusion, or approximate similarity. When query functionality is available, it often relies on rigid form-based inputs or formal query languages that are difficult for non-expert users. Furthermore, many tools are built for lightweight ontologies with simple taxonomies and do not scale effectively to large, expressive ontologies with rich relational semantics. These limitations motivate the development of new tools that support intuitive and interactive construction of complex queries and enable semantic search across expressive, real-world ontologies.

\section{Technical Background} \label{sec:preliminaries}

This section provides the necessary background and preliminaries. We first review the basics of ontologies in Section~\ref{sec:review-ontologies}, followed by an overview of fuzzy ontologies in Section~\ref{sec:dl-fuzzy}.

\subsection{Ontologies: Syntax and Semantics} \label{sec:review-ontologies}

In this work, we adopt the formalism of \textit{Description Logics (DLs)}~\cite{baader2003description} to describe the syntax and semantics of ontologies, which is necessary for presenting our ontology embeddings. DLs are a family of logic-based languages that provide a formal, interpretable syntax and well-defined semantics for representing ontologies. They are widely used and form the foundation of popular ontology languages such as OWL~\cite{mcguinness2004owl}, which combines DLs with the RDF data model~\cite{klyne2006resource}. Although our approach is not limited to DLs, we use DL notation to review fuzzy ontologies and their interpretations in order to clearly define our embeddings.

Formally, the syntax of a DL ontology \(\ontology = (\Sigma, \mathcal{K})\) consists of a \emph{signature} \(\Sigma = (\concepts, \roles, \inds)\), which defines the vocabulary of concept names, role (relationship) names, and individual names, and a \emph{knowledge base} \(\mathcal{K} = (\tbox, \abox)\) containing two types of axioms. The \textit{TBox} (\(\tbox\)) specifies relationships between concepts, including subsumption axioms (e.g., \(C \sqsubseteq D\)), equivalence axioms (\(C \equiv D\)), and disjointness axioms (\(C \sqcap D \sqsubseteq \bot\)). The \textit{ABox} (\(\abox\)) contains assertions about individuals, such as \(a : C\), indicating that individual \(a\) is an instance of concept \(C\), or \(R(a, b)\), stating that individuals \(a\) and \(b\) are related via the role \(R\). The semantics of an ontology is defined by an \textit{interpretation} \(\mathcal{I} = (\Delta^\mathcal{I}, \cdot^\mathcal{I})\), where \(\Delta^\mathcal{I}\) is a non-empty domain and \(\cdot^\mathcal{I}\) is an interpretation function. This function maps each concept \(C\) to a subset \(C^\mathcal{I} \subseteq \Delta^\mathcal{I}\), each role \(R\) to a binary relation \(R^\mathcal{I} \subseteq \Delta^\mathcal{I} \times \Delta^\mathcal{I}\), and each individual \(a\) to an element \(a^\mathcal{I} \in \Delta^\mathcal{I}\).

Complex (or defined) concepts are constructed using logical constructors such as conjunction (\(C \sqcap D\)), disjunction (\(C \sqcup D\)), and negation (\(\lnot C\)), which correspond to standard set operations in the domain. Specifically, the interpretation of a conjunction is given by \((C \sqcap D)^\mathcal{I} = C^\mathcal{I} \cap D^\mathcal{I}\), the interpretation of a disjunction is \((C \sqcup D)^\mathcal{I} = C^\mathcal{I} \cup D^\mathcal{I}\), and the interpretation of a negation is \((\lnot C)^\mathcal{I} = \Delta^\mathcal{I} \setminus C^\mathcal{I}\).

An ontology is said to be \textit{consistent} if all of its axioms can be satisfied simultaneously in some interpretation. Consistency is a basic requirement for reasoning and querying, as inconsistencies can make the ontology logically meaningless or prevent useful inference. A concept is \textit{satisfiable} if it has at least one instance in some interpretation; otherwise, it is considered logically contradictory. These notions provide the foundation for standard reasoning tasks such as \textit{subsumption checking}, which determines whether \(C \sqsubseteq D\) holds in all interpretations (i.e., whether every instance of \(C\) is also an instance of \(D\)), and \textit{instance checking}, which verifies whether an individual \(a\) necessarily belongs to a concept \(C\) across all models of the ontology. While such reasoning tasks support classification and consistency checking, they are typically handled internally by DL reasoners.

In our setting, we are more concerned with querying ontologies to retrieve individuals or concepts that match a user's information needs. Common query types include conjunctive queries (CQs), unions of conjunctive queries (UCQs), and recursive extensions that support reachability-like conditions. DL query languages such as SPARQL-DL~\cite{sirin2007sparql}, nRQL~\cite{haarslev2004nrql}, and EQL-Lite~\cite{calvanese2007eql} combine logical reasoning with structured query syntax. However, these languages require familiarity with formal syntax and detailed knowledge of the ontology’s structure, and their evaluation is often computationally expensive. This limits their suitability for users seeking intuitive, flexible, and interactive exploration.

\subsection{Fuzzy Ontologies} \label{sec:dl-fuzzy}

Fuzzy ontologies extend classical ontologies by incorporating fuzzy logic, a many-valued logic in which truth values range continuously between 0 and 1. This enables modeling concepts and relationships that are inherently vague or uncertain. In contrast to crisp ontologies, where an individual either belongs to a concept or not, fuzzy ontologies allow individuals to belong to a concept with a certain degree of membership~\cite{bobillo2011fuzzy, tho2006automatic}. For example, a patient might belong to the concept \textit{Diabetic} with degree 0.8, reflecting partial evidence or uncertain data.

Fuzzy Description Logics (fuzzy DLs) extend classical DLs with fuzzy semantics, enabling formal reasoning over fuzzy sets~\cite{straccia1998fuzzy, hajek2005making, garci2010fuzzy}. As in the classical case, a fuzzy ontology \(\ontology = (\Sigma, \mathcal{K})\) has a signature \(\Sigma = (\concepts, \roles, \inds)\) and a knowledge base \(\mathcal{K} = (\tbox, \abox)\), but its interpretation function changes. Some fuzzy DL variants also introduce optional syntactic extensions, such as explicit degree annotations or fuzzy modifiers in axioms or assertions. 

An interpretation \(\mathcal{I} = (\Delta^\mathcal{I}, \cdot^\mathcal{I})\) now maps each concept \(C\) to a \emph{membership function} \(C^\mathcal{I}: \Delta^\mathcal{I} \to [0,1]\), where \(C^\mathcal{I}(x)\) gives the degree to which individual \(x\) belongs to \(C\). Similarly, roles are interpreted as fuzzy binary relations \(R^\mathcal{I}: \Delta^\mathcal{I} \times \Delta^\mathcal{I} \to [0,1]\).

Complex concepts are interpreted using fuzzy operators. Given fuzzy membership functions \(\mu_C\) and \(\mu_D\) for concepts \(C\) and \(D\), conjunction is defined as \((C \sqcap D)^\mathcal{I}(x) = \tnorm(\mu_C(x), \mu_D(x))\), disjunction as \((C \sqcup D)^\mathcal{I}(x) = \tconorm(\mu_C(x), \mu_D(x))\), and negation as \((\lnot C)^\mathcal{I}(x) = \fneg(\mu_C(x))\), where \(\tnorm\) is a continuous t-norm (e.g., product, Gödel, or Łukasiewicz), \(\tconorm\) is its corresponding t-conorm, and \(\fneg\) is a negation function~\cite{esteva2000residuated, garci2010fuzzy}. These operators preserve the semantics of fuzzy set theory and allow the construction of defined concepts with graded membership.

Fuzzy DL also generalizes the notion of axiom satisfaction. A concept assertion \(\langle x : C \geq \alpha \rangle\) is satisfied if the membership degree of \(x\) in \(C\) is at least \(\alpha\), i.e., \(\mu_C(x) \geq \alpha\). A subsumption axiom \(\langle C \sqsubseteq D \geq \alpha \rangle\) is satisfied if, for all individuals in the domain, the degree to which \(C\) implies \(D\) is at least \(\alpha\). This implication is computed using the residuum of the chosen t-norm, a standard fuzzy logic operation that captures graded entailment. An ontology is consistent to degree \(\alpha\) if all its axioms are satisfied to at least that degree.

In this work, we adopt a fuzzy Description Logic framework that supports partial membership in defined concepts and enables fuzzy reasoning over user-defined queries. Our implementation is based on fuzzy ontology reasoners and uses continuous t-norms to define fuzzy operations. This allows for flexible, approximate, and semantically grounded matching of complex concepts in expressive ontologies.

\section{Ontology Querying using Fuzzy Ontology Embeddings}
\label{sec:main}

We now present our fuzzy ontology embedding approach and describe how it enables flexible, semantically meaningful queries over large ontologies.

Let \(\ontology\) be an ontology and let \(\mathcal{I} = (\Delta^\mathcal{I}, \cdot^\mathcal{I})\) be a fuzzy interpretation of \(\ontology\), where \(\Delta^\mathcal{I} = \{x_1, \ldots, x_d\}\) is a fixed set of domain elements and \(\cdot^\mathcal{I}\) assigns to each concept \(C \in \concepts\) a fuzzy membership function \(\mu_C^\mathcal{I} : \Delta^\mathcal{I} \to [0,1]\). The embedding of a primitive concept \(C\) is defined as:%
\begin{align}
\hspace{3.5cm}\mathbf{v}_C = [\mu_C^\mathcal{I}(x_1), \mu_C^\mathcal{I}(x_2), \ldots, \mu_C^\mathcal{I}(x_d)] \in [0,1]^d.    
\end{align}

Essentially, each domain element \(x_i\) acts as a point in the ontology's semantic space. For each concept \(C\), the index corresponding to \(x_i\) is a score of how 'close' or 'inside' an element is to \(C\). The concept embeddings are vectors of these scores, positioning each concept in the semantic space relative to the domain elements. This allows us to represent each concept not as arbitrary points, but as regions capturing a semantic profile. Given a set of domain elements representing a hierarchical ontology, a root concept would have high membership scores to all the domain elements. The embedding of this root would reflect that the concept contains everything in the ontology's semantic space. Whereas, a leaf concept would likely only have sparse membership to a few or a single domain element and the leaf's embedding would represent this.

Embeddings for composite concepts are computed on demand using fuzzy logic operations applied element-wise to primitive embeddings. Let \( \bar{\theta} \), \( \bar{\kappa} \), and \( \bar{\fneg} \) denote the element-wise versions of a chosen t-norm, t-conorm, and negation operator, respectively. For example, the embedding of \( C_1 \sqcap C_2 \) is \(\mathbf{v}_{C_1 \sqcap C_2} = \bar{\theta}(\mathbf{v}_{C_1}, \mathbf{v}_{C_2})\), where \( \bar{\theta} \) could be, for instance, the product t-norm \( \bar{\theta}(a, b) = a \cdot b \). Other common choices include the Gödel and Łukasiewicz t-norms. Disjunctions and negations are handled analogously using \( \bar{\kappa} \) and \( \bar{\fneg} \), such as \( \bar{\kappa}(a, b) = a + b - a \cdot b \) and \( \bar{\fneg}(a) = 1 - a \).

A query \(Q\) is a composite concept that is defined compositional using primitive concepts and other composite concepts using fuzzy operators---conjunction (\( \sqcap \)), disjunction (\( \sqcup \)), and negation (\( \lnot \))---without requiring these concepts to be explicitly present in the ontology. Each such query \( Q \) is represented by a vector \( f_m(Q) \in [0,1]^d \), computed recursively from its structure using the fuzzy operators. Answering the query is done by retrieving similar primitive concepts. To retrieve relevant concepts, the system compares the embedding \( f_m(Q) \) of a user-defined query with all primitive concept embeddings using cosine similarity and returns the top-\( k \) most similar concepts. This enables exploratory, partial, and approximate matches that reflect semantic similarity rather than exact equivalence. Once a fuzzy interpretation is computed, the membership vectors for all primitive concepts can be precomputed and stored in a vector database. At query time, composite concept embeddings are computed dynamically using fuzzy operations, and similarity search is performed efficiently over the stored embeddings. This design enables fast, scalable, and semantically grounded concept-based retrieval without the need for repeated reasoning or re-embedding.

The quality of embeddings and the query results depend critically on the fuzzy interpretation \( \mathcal{I} \) used to assign membership degrees to domain elements. This interpretation must approximately satisfy the ontology's axioms, ensuring semantic coherence. However, axiomatic satisfaction alone may not guarantee high-quality embeddings. For instance, embeddings should ideally reflect concept hierarchy, assign distinct vectors to semantically distinct concepts, and group semantically similar individuals with similar membership degrees. These additional criteria can be encouraged by integrating auxiliary axioms or constraints---e.g., enforcing similarity among sibling concepts or ensuring disjointness of non-overlapping concepts---even if such constraints are not part of the original ontology.

Our approach is agnostic to the specific fuzzy DL reasoner used to derive \( \mathcal{I} \), as long as it yields consistent and interpretable membership functions. Several off-the-shelf reasoners can be used. FALCON~\cite{tang2022falcon} learns fuzzy interpretations via sampling and optimization. Users can specify the sample size, which defines the number of domain elements \( |\Delta^\mathcal{I}| \) in the interpretation and thus determines the embedding dimension \( d \). This flexibility allows users to trade off between precision and computational cost. fuzzyDL~\cite{bobillo2008fuzzyDL} is a tableau-based reasoner that computes crisp or fuzzy satisfiability of axioms. It does not expose direct control over the size of the domain used for interpretation, making it less flexible for applications that require embedding generation over a fixed-size universe. DeLorean~\cite{penaloza2010deorean} approximates fuzzy reasoning using reduction techniques that map fuzzy ontologies into classical ones. It supports reasoning over various t-norms and fuzzy operators but similarly lacks user control over the interpretation domain. 
Among these, FALCON offers the most direct way to control the embedding space, as the number of sampled individuals directly determines the embedding dimensionality. This tunable design is well-suited for downstream applications that require balancing interpretability, query speed, and representational capacity.

Our querying approach and fuzzy ontology embeddings offer several advantages that make it well-suited for semantically grounded, scalable, and interactive ontology exploration:%
\begin{itemize}
    \item {\em Semantically grounded embeddings:} The embeddings are derived from a fuzzy interpretation \( \mathcal{I} \) that satisfies the ontology’s axioms, ensuring that the resulting vectors reflect the intended semantics of the ontology. Unlike many prior embeddings that rely on taxonomic structures or shallow graph features, our method supports ontologies with expressive axioms, provided they are supported by the fuzzy DL reasoner. This allows meaningful semantic representation even in complex, non-hierarchical ontologies.

    \item {\em Compositionality and real-time querying:} The embeddings are compositional: primitive concept vectors are precomputed once and stored offline, while composite concepts are constructed at runtime using efficient vectorized fuzzy operations. This enables real-time querying without retraining or re-embedding, in contrast to neural or graph-based approaches that require recomputation when new concepts are introduced.

    \item {\em Tunable embedding quality and size:} The embedding dimensionality and fidelity are determined by the size of the interpretation domain \( |\Delta^\mathcal{I}| \), which can be controlled via parameters exposed by the fuzzy reasoner (e.g., the sample size in FALCON~\cite{tang2022falcon}). This provides a flexible trade-off between representational richness and computational efficiency.

    \item {\em Scalable and optimized retrieval:} Since all concept representations are vectors, similarity search over primitive concepts can be delegated to optimized vector databases (e.g., FAISS~\cite{johnson2019faiss}, Qdrant, or Chroma~\cite{chroma2025}). These systems support fast top-\( k \) retrieval and efficient indexing, enabling responsive and semantics-aware retrieval for exploratory query interfaces.
\end{itemize}

\subsection{\(\alpha\)-Embeddings for Hierarchical Ontologies}
\label{sec:hierarchical-interpretation}

In many practical domains such as biomedicine, ontologies take the form of hierarchies or trees, where each concept has a parent concept (possibly more), some number of children, and there are no cycles (e.g., HPO~\cite{human2024nucl}, PO~\cite{jaiswal2005plant}). For such ontologies, we can construct a synthetic but semantically meaningful fuzzy interpretation \( \mathcal{I} \) that captures hierarchical relationships and allows us to generate expressive embeddings efficiently. We refer to these synthetic embeddings as \(\alpha\)-embeddings.

We assume the ontology \( \ontology \) forms a tree-structured taxonomy, where all paths lead from leaves to the root through unique parent-child links. Let \( \Delta^\mathcal{I} = \{x_1, \ldots, x_d\} \) be the interpretation domain constructed iteratively as follows:%
\begin{enumerate}
    \item At each step, a new domain element \( x_i \) is added. 
    \item A leaf concept \( C \in \concepts \) is chosen uniformly at random.
    \item The membership of \( x_i \) to \( C \) is set to 1: \( \mu_C^\mathcal{I}(x_i) = 1 \).
    \item For every other leaf concept \( C' \), we compute the graph distance \( d(C, C') \), defined as the number of edges to their lowest common ancestor. We assign decaying membership using a fading parameter \( \alpha \in (0,1) \): 
    \begin{equation*}
        \mu_{C'}^\mathcal{I}(x_i) = \alpha^{d(C, C')}.
    \end{equation*}
    \item For any internal (non-leaf) concept \( C_p \) with children \( C_1, \ldots, C_n \), the membership is defined recursively using the fuzzy union aggregation (e.g., probabilistic sum \( \kappa \)):
    \begin{equation*}      
        \mu_{C_p}^\mathcal{I}(x_i) = \kappa\big(\mu_{C_1}^\mathcal{I}(x_i), \ldots, \mu_{C_n}^\mathcal{I}(x_i)\big).
    \end{equation*}
\end{enumerate}

This process generates a coherent fuzzy interpretation over a hierarchical ontology. By construction, the subsumption axioms \( C_j \sqsubseteq C_p \) are satisfied in the fuzzy sense, since the membership degree of any domain element to a parent concept \( C_p \) is at least as large as that to its children, due to the properties of the fuzzy union. Thus, the generated \( \mathcal{I} \) satisfies all taxonomic axioms of the ontology.

This method provides an effective and interpretable way to construct embeddings from hierarchical ontologies, capturing both local concept similarity and the global structure. It also allows control over smoothness and generalization via the decay parameter \( \alpha \), making it a practical tool for building fuzzy ontology embeddings in common tree-structured settings.
\section{The \floqe\ System} \label{sec:floqe}

In this section, we present our prototype system, {\em \floqe}. Section~\ref{sec:floqe-architecture} describes its architecture and important implementation details. Section~\ref{sec:floqe-back} further discusses the back-end server functionality. Section~\ref{sec:floqe-front} does the same for the front-end interface. Lastly, Section~\ref{sec:floqe-sce} explores the prior Example~\ref{ex:intro} in greater depth as a usage scenarios that show how \floqe\ supports ontology exploration and search through visual query building and approximate reasoning.

\subsection{\floqe\ Architecture and Implementation} \label{sec:floqe-architecture}

\begin{figure}[htb]
    \centering
    \includegraphics[width=1.0\linewidth]{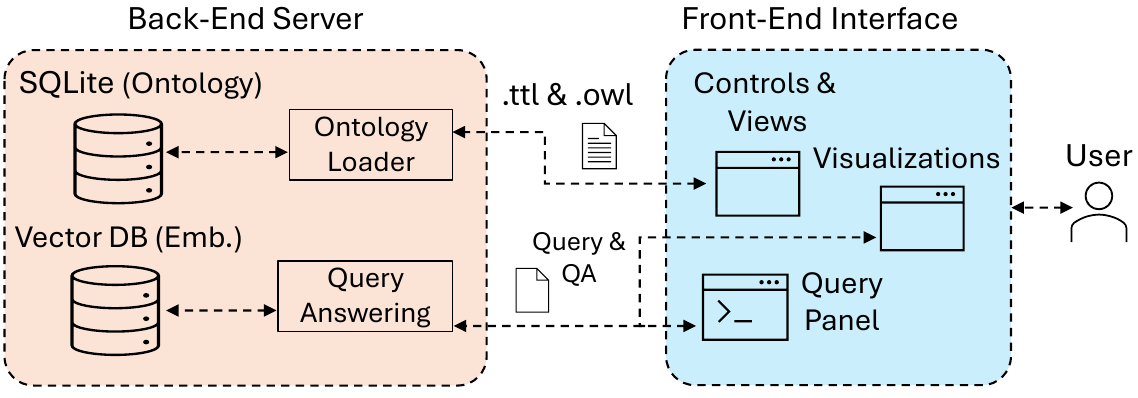}
    \caption{An overview of the architecture of \floqe}
    \label{fig:system-overview}
\end{figure}

\floqe\ is a web-based system designed for ontology exploration. Figure~\ref{fig:system-overview} shows its overall architecture. The front-end is built using HTML5, CSS, and JavaScript, with D3.js (version 7) for interactive visualizations. The back-end is implemented in Python and communicates with the front-end via a FLASK API.

The system workflow begins with the user uploading ontology data in standard formats such as \texttt{.owl} or \texttt{.ttl}. The back-end then processes and stores the ontology, making it available for the front-end. Once loaded in the front-end, users can explore the ontology through interactive views and visualizations. This exploration is supported by panels and controls that allow users to manipulate views, adjust parameters, and construct queries. As users interact with the system, \floqe's front-end dynamically retrieves relevant ontology data from the back-end.

While the front-end handles user interaction and visualization, all computations---including query evaluation and similarity search---are performed on the back-end. Embeddings for primitive concepts are stored in a Chroma vector database, as described in Section~\ref{sec:main}. This enables fast query resolution, especially for ontologies with thousands of concepts and/or large embeddings.

The remainder of this section describes the back-end (Section~\ref{sec:floqe-back}) and front-end (Section~\ref{sec:floqe-front}) in more detail, following the architecture shown in Figure~\ref{fig:system-overview}.

\subsection{Back-End Server} \label{sec:floqe-back}

The back-end consists of four main components: the ontology loader, ontology database (SQLite), vector (embedding) database, and the query answering component. The ontology loader handles uploaded ontologies---pre-uploaded or submitted via the front-end. It parses ontology files and stores their contents in an ontology database using the OwlReady2 python library~\cite{lamy2017owlready}. During this process, additional metadata and statistics---such as subtree sizes and number of children---are computed and stored. The vector database stores the embeddings for each primitive concept which are generated using either a fuzzy ontology reasoner or our $\alpha$-embeddings as we explained in Section~\ref{sec:hierarchical-interpretation}. Just like for ontologies, users can submit embedding vectors or use pre-uploaded ones. The vector database is separate from the ontology database to perform quick similarity calculation and top-\(k\) retrieval. Our implementation uses Chroma for its vector database. The query answering component computes embeddings for user-defined composite concepts using the embeddings of primitive concepts from the vector database. The vector database uses these composite concept embeddings to retrieve similar primitive concepts. A FLASK server API handles communication between the back- and front-end. 

\subsection{Front-End Interface} \label{sec:floqe-front}
\begin{figure}[htb]
    \centering
    \setlength{\fboxsep}{0.5pt}\fbox{\includegraphics[width=0.98\linewidth]{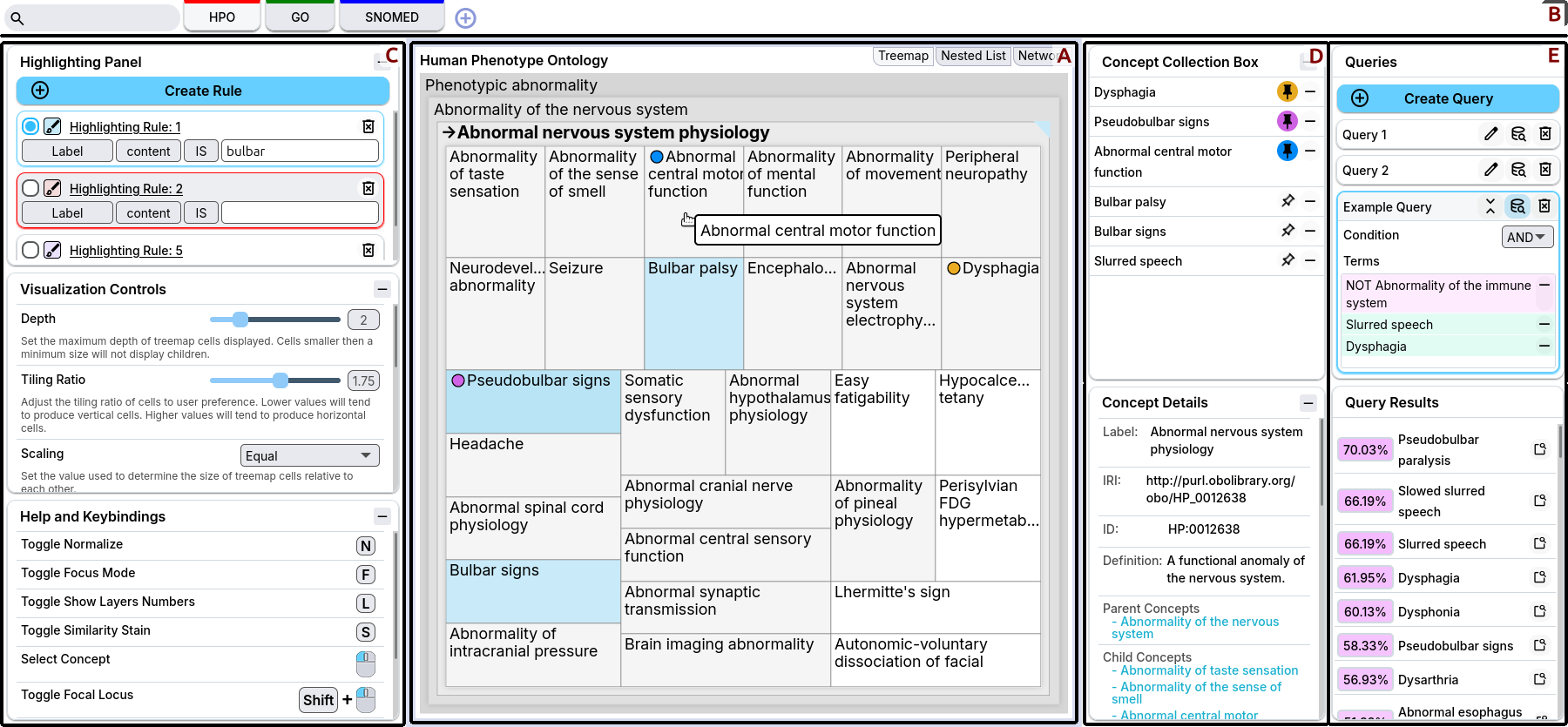}}
    \caption{An overview of \floqe. A) The primary visualizations, B) the top header, C) the left ontology building section, and D) the right query section.}
    \label{fig:overview}
\end{figure}

The front-end is composed of several regions consisting of panels and controls that support interactive ontology exploration. Figure~\ref{fig:overview} shows an overview of the front-end interface as seen by the user. The primary visualization takes up the central region of the interface and is the user's primary means of navigating the ontology (Figure~\ref{fig:overview}.A). Users can change between different visualizations---treemps, nested lists, and network graphs---in this central region. The second region (Figure~\ref{fig:overview}.B), the top header, provides quick access for concept search, switching between loaded ontologies, and loading ontologies. Users can manipulate the behavior and control the presentation using the visualization controls in the leftmost region (Figure~\ref{fig:overview}.C). The first of the rightmost regions (Figure~\ref{fig:overview}.D) is the concept panels where users can store concepts for future use and look at concept details. The second rightmost region (Figure~\ref{fig:overview}.E) is the query building panels where users define queries and examine their results. \floqe’s front-end is compatible with all major web browsers.

Below, we describe the main components of the front-end interface in more detail.

\subsubsection{Primary Visualizations}

\floqe’s central region contains the primary visualization (Figure~\ref{fig:overview}.A). Here, users see a representation of the ontology to aid them in exploration and understanding. Each component of \floqe\ serves to modify, control, and navigate the primary visualization, with the back-end providing any information needed by the primary visualization. The primary visualization can take the form of any suitable representation of the ontology. 
Visualization should meet the following criteria:
\begin{itemize}
    \item \textit{Concept Distinguishability}. Each concept in the visualization must be distinct and individually identifiable, at a reasonable scale. The density of concepts must be low enough that users can distinguish them. As well, concepts must have some clear identifier---often this is a label.
    
    \item \textit{Consistent}. The visualization should represent concepts and relationships in a consistent and identifiable manner. 
    
    \item \textit{Dynamic and Interactive}. The visualization should react to users’ actions. Static representations may be sufficient for small and simple datasets, but ontologies are often neither small nor simple. Dynamic visualizations adapt to users needs and can change to show the ontology information that they are interested in---especially when presenting the entire ontology is impractical. 
    
    \item \textit{Property-revealing}. The visualization should present secondary information and metadata of concepts. Features such as concept descriptions, depth, subtree size, and metadata contain information necessary to understand the content of ontologies. Users may be unfamiliar with concept labels and require secondary information to understand an ontology.
     
    \item \textit{Space-efficient}. The visualization should utilize the available space fully. Some ontologies have a large amount of content and visualizations that make poor use of space are either sparse or dense to the point of uselessness. 
    
    \item \textit{Structure-revealing}. The visualization must showcase the primary semantic structure of the ontology. In most cases, this is the concept subsumption hierarchy. One of the primary purposes of ontologies is to express how concepts relate to one another. This information is critical to users’ understanding of ontologies.
\end{itemize}

Classical visualizations such as nested lists, icicle charts, network graphs, and treemaps only partially meet these requirements. In fact, no single static visualization can satisfy every requirement. Network graphs become unreadable with large numbers of concepts, nested lists waste space, and treemaps struggle to convey secondary information. Our approach addresses these limitations by combining the primary visualization with secondary panels for additional details and utilizing dynamic interactive visualizations. To illustrate how \floqe\ meets these requirements, we describe the nested-treemap version of the primary visualization.

\begin{figure}[htb]
    \centering
    \setlength{\fboxsep}{0pt}\fbox{\includegraphics[width=\linewidth]{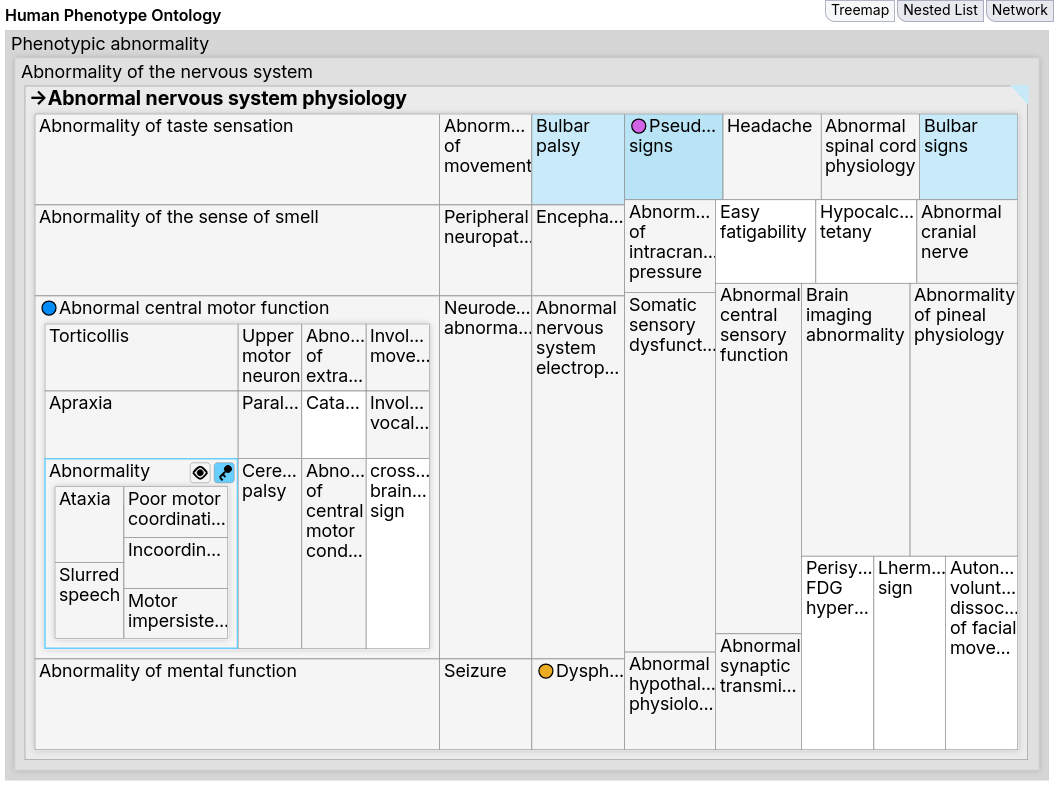}}
    \caption{The primary visualization set to the treemap view. The \nit{Abnormal nervous system physiology} concept is selected and focus mode is enabled with \nit{Abnormality of coordination} set as locus.}
    \label{fig:treemap-0}
\end{figure}

Nested-treemaps represent concepts as rectangles, place sibling concepts adjacent to each other, and nest child concepts inside their parent's area to indicate hierarchy. They are well-suited for ontology visualization because they clearly distinguish concepts, are consistent, space-efficient, and structure-revealing. Extending nested-treemaps to be dynamic and interactive allows for the display of secondary properties and addresses the problem of excessive concept density. Figure~\ref{fig:treemap-0} shows \floqe’s nested-treemap (hereafter, \textit{the treemap}), which displays concepts within an adjustable distance of a selected concept. As users navigate, the treemap updates to show related concepts and integrates with other \floqe\ front-end components to support exploration.

Selecting a concept in the treemap refocuses the visualization onto it, allowing users to drill down into its subtree or navigate upward to parent concepts. Users can also drag concepts to other components and pin them for later reference. Pinning creates visual marks that aid in orientation by acting as signposts. These marks help identify when a concept appears in multiple subtrees. For example, in HPO the \nit{Cardiac valve calcification} concept is a child of both \nit{Cardiovascular calcification} and \nit{Abnormal heart morphology}; when pinned, it becomes more noticeable in both locations, indicating structural overlap. Without such aids, users would need to manually read labels to detect these cases. Once users have collected concepts of interest, they can create queries using the Query Building Panels to guide further exploration.

Users may want to focus on a concept while retaining its context within the ontology. Simply increasing depth causes two issues. One, deeper child concepts receive minimal space, forcing users to mouse over for tooltips to identify them. Two, densely populated subtrees create visual clutter that overwhelms users. Selecting the concept alone does not solve this, as it refocuses the visualization and removes the surrounding context.

To address this, we implemented a toggle for a \textit{focus mode} that enables on-the-fly visual distortions. During focus mode, when users mouse over a concept, \floqe\ applies a discrete Cartesian distortion (often referred to as a fisheye distortion or lens), enlarging its area while preserving relative position. Within the expanded area, the subtree extends, showing the focused concept's children; users can drill down further by mousing over these until no more space is available. If users want to explore multiple subtrees simultaneously they can lock a concept as a locus and the distortion will maintain multiple focal points. Moreover, locked loci can be transformed into a different visualization. For example, a concept cell can be converted to a node network graph or a panel showing class details. In essence, this technique directly addresses the problem of maintaining context while refocusing.

Figure~\ref{fig:treemap-0} shows an example of focus mode in use. With depth set to two and \nit{Abnormal nervous system physiology} selected, no concepts other than its children would be visible. Yet, with focus mode turned on, hovering over \nit{Abnormal central motor function} fills the subtree with child concepts. Afterwards, when the user mouses over the child \nit{Abnormality of coordination}, the subtree is further extended to show this concept’s children (e.g., \nit{Slurred speech}). This can continue until a leaf node is reached or space runs out, but the user decided this depth was sufficient and locked \nit{Abnormality of coordination} as a locus. Focus mode thus offers a dynamic and space-efficient way to explore deeper while maintaining orientation. Furthermore, dynamic distortions can be applied to other kinds of visualizations, such as network graphs and charts.

Users adjust the primary visualization through the Visualization Controls (Section~\ref{sec:floqe-v-controls}). For nested-treemaps, users can adjust depth, the label content (e.g., depth numbers), tiling methods, concept visibility, and highlighting. Such controls are essential, as different ontologies are best viewed with different settings. For instance, in HPO, \nit{Abnormality of the musculoskeletal system} dominates the space when scaling is proportional to subtree size, making the smaller \nit{Abnormality of the voice} nearly invisible. Equal scaling makes both visible, reducing the chance of missing concepts. Furthermore, users can toggle such options to see how the visualization changes---turning equal scaling on and off shows users which concepts take up the majority of an ontology while quickly returning to a view suitable for navigation. These adjustments enable richer exploration without restricting users to suboptimal visualizations.

\floqe’s primary visualization, in conjunction with other front-end panels, provides a spatial overview of ontology structure and semantics. By supporting dynamic, interactive navigation, it helps users quickly form mental models and identify key regions of interest---an essential capability for working with complex knowledge domains where manual inspection would be slow and overwhelming. Ultimately, the primary visualization serves as the main interface through which users interact with and explore ontologies.

\subsubsection{Header Bar}
\begin{figure}[htb]
    \centering
    \setlength{\fboxsep}{0pt}\fbox{\includegraphics[width=0.45\linewidth]{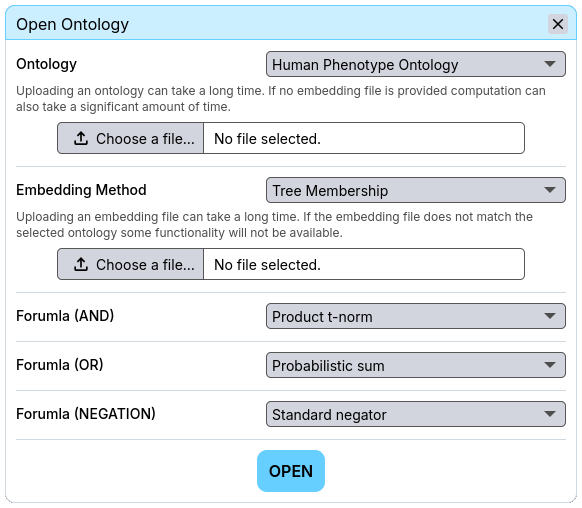}}
    \caption{The ontology selection menu. A number of options allow users to manipulate the system's behavior.}
    \label{fig:startup-view}
\end{figure}

\floqe's header bar (Figure~\ref{fig:overview}.B) contains three interactive elements: a search bar for concept lookup, tabs for switching between ontology instances, and a plus button to open the ontology selection panel (Figure~\ref{fig:startup-view}). These allow users to load ontologies, switch among them, and perform quick keyword searches.

The search bar allows users to find concepts by label and navigate directly to them. The system matches the input text against concept labels in the current ontology, enabling rapid navigation for users familiar with the domain and quick verification of concept existence. For example, in a biomedical ontology, searching for \nit{Heart} confirms whether it is present and suggests related results (e.g., \nit{Heart block}) for discovery. While not intended as the primary navigation method, it is useful for repositioning to known landmark concepts.

The next header element---tabs---allows users to switch between multiple loaded ontologies. Each retains its own settings, highlight rules, pinned concepts, queries, and visualizations. Tabs support parallel exploration of related ontologies, comparison of different embeddings and fuzzy operators, or exploration of separate regions within a large ontology without losing placement. Users can organize queries, create dedicated highlight rules, and compare configurations across instances. New tabs are added via the ontology selection menu.

The plus button opens the ontology selection panel (Figure~\ref{fig:startup-view}), where users can choose from preprocessed ontologies or upload their own. Uploaded files are processed by the back-end, which stores each ontology and its embedding in separate temporary databases; preprocessed ontologies use static databases. Embeddings can be uploaded or selected from pre-made options to evaluate effectiveness. Users can also configure fuzzy logic operators for query resolution. Depending on the fuzzy DL reasoner used, aligning an embedding with the matching fuzzy operators may be necessary. The provided product, Gödel, and Łukasiewicz t-norms cover most cases---often functioning well even when not perfectly matched to the reasoner.

\subsubsection{Visualization Controls} \label{sec:floqe-v-controls}
\begin{figure}[htb]
    \centering
    \begin{subfigure}[t]{.48\textwidth}
        \centering
        \captionsetup{width=.9\linewidth}
        \setlength{\fboxsep}{0pt}\fbox{\includegraphics[width=0.9\linewidth]{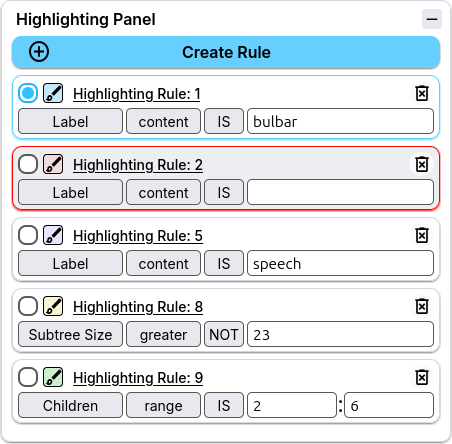}}
        \caption{The highlight panel showing multiple rules for staining the primary visualization.}
        \label{fig:highlight-control}
    \end{subfigure}
    \begin{subfigure}[t]{.48\textwidth}
        \centering
        \captionsetup{width=.9\linewidth}
        \setlength{\fboxsep}{0pt}\fbox{\includegraphics[width=0.9\linewidth]{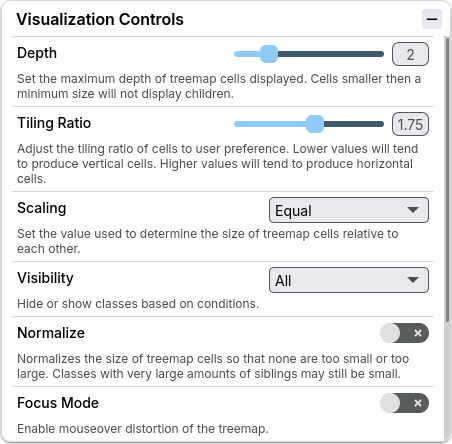}}
        \caption{The visualization control panel for a treemap visualization.}
        \label{fig:visualization-control}
    \end{subfigure}
    \caption{The highlight panel and visualization controls of the leftmost region of \floqe. Used for manipulating the primary visualization.}
    \label{fig:left-overview}
\end{figure}

The left region of \floqe\ contains controls and panels for adjusting the primary visualization (Figure~\ref{fig:overview}.C). To illustrate, Figure~\ref{fig:left-overview} shows the panels for a nested-treemap visualization. This region has three components: the highlight panel for applying color-based rules (Figure~\ref{fig:highlight-control}), the visualization control panel for adjusting the primary visualization's behavior (Figure~\ref{fig:visualization-control}), and a help panel explaining system interaction. These controls let users tailor the visualization to their needs, as static parameters are often insufficient. For example, a user on a small monitor may require a more horizontal treemap layout to read labels without hovering.

The highlight panel (Figure~\ref{fig:highlight-control}) allows users to annotate the primary visualization by creating rules that assign colors to concepts. When a concept satisfies multiple rules, colors are blended. Rules can filter by label, subtree size, number of children, or other criteria; string-based filters may require exact or substring matches, while numerical filters can specify ranges, minimums, maximums, or equality. Rules can also be negated or toggled on and off.

For example, a rule may color concepts blue if their labels contain \textit{bulbar} (Figure~\ref{fig:highlight-control}, first rule), making concepts such as \nit{Bulbar Palsy} and \nit{Pseudobulbar signs} stand out. Highlighting helps users quickly identify concepts of interest to form landmark and neighborhood knowledge. Isolated colored concepts may indicate uniqueness, while clusters can suggest similarity. Users can also choose to display only highlighted concepts, which is especially useful in dense visualizations such as treemaps or network graphs. Highlighting thus provides essential visual cues for navigation and can be extended to support ontology-specific rules when needed.

The visualization control panel configures the behavior of the primary visualization, with available options varying by visualization type. For example, Figure~\ref{fig:visualization-control} shows the controls for a nested-treemap, which include:%
\begin{itemize}
    \item \emph{Depth.} Maximum displayed depth from the selected concept. Higher values pack more content for rapid traversal; lower values improve clarity.
    \item \emph{Tiling Ratio.} Controls the horizontal–vertical balance of treemap cells. Screen size and ontology structure affect which ratios best preserve readability; for example, a strongly horizontal layout aids long labels but may have insufficient vertical space when many concepts are present.
    \item \emph{Scaling.} Determines concept size. Scaling by subtree size or number of children emphasizes dominant concepts, while equal scaling reveals smaller ones that might otherwise be hidden.
    \item \emph{Visibility.} Controls which concepts are shown, often based on highlighting rules, allowing irrelevant concepts to be hidden.
    \item \emph{Normalization.} Adjusts scaling to make small concepts more visible and large ones less dominant.
    \item \emph{Focus Mode.} Enables mouse over distortion, dynamically enlarging concepts of interest regardless of ontology structure.
    \item \emph{Showing Layer Numbers.} Annotates concepts with their depth, useful in visualizations (e.g., network graphs) where depth is hard to gauge.
    \item \emph{Toggling Similarity Stain.} Colors concepts based on query results, quickly showing which areas of the ontology match.
\end{itemize}
Some settings are universal, while others are visualization-specific. These controls let users adapt the visualization to their needs, as default settings are rarely optimal for all ontologies.

The final panel—help and keybindings—provides usage guidance and can be hidden if not needed. While a full interactive tutorial would be ideal, it is beyond the scope of this prototype.

\subsubsection{Concept Panels}
The first section of the right region of \floqe\ contains the concept panels (Figure~\ref{fig:overview}.D), which serve two purposes: organizing the collection of concepts for later use and displaying concept details. As users explore ontologies, they may find concepts worth revisiting or using in queries. These concepts can be scattered across the ontology, making them inconvenient to access without a collection. Details are also needed to confirm relevance. For example, in HPO, a user studying skull-related conditions may collect both \nit{Abnormal cranial nerve physiology} and \nit{Abnormality of the head}, despite their distance in the hierarchy. Storing them in the collection allows quick navigation between them. Overall, the concept panels aids in navigation and query formulation.

\begin{figure}[htb]
    \centering
    \begin{subfigure}[t]{.30\textwidth}
        \centering
        \captionsetup{width=.9\linewidth}
        \setlength{\fboxsep}{0pt}\fbox{\includegraphics[width=0.9\linewidth]{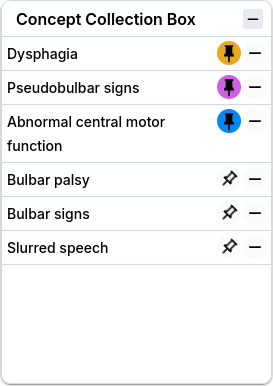}}
        \caption{The concept collection panel with some pinned concepts.}
        \label{fig:concept-collection}
    \end{subfigure}
    \begin{subfigure}[t]{.30\textwidth}
        \centering
        \captionsetup{width=.9\linewidth}
        \setlength{\fboxsep}{0pt}\fbox{\includegraphics[width=0.9\linewidth]{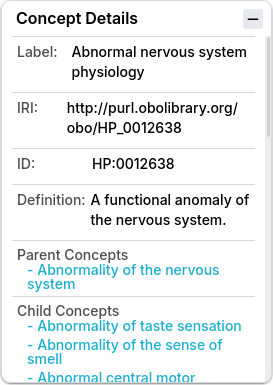}}
        \caption{The details view panel showing the \nit{Abnormality of the eye} concept}
        \label{fig:concept-details}
    \end{subfigure}
    \caption{The concept panel of the rightmost region of \floqe. Used for storing and analyzing concepts.}
    \label{fig:right-overview-0}
\end{figure}

The collection panel (Figure~\ref{fig:concept-collection}) stores user-selected concepts for easy access. Users can add concepts to the panel by drag-and-drop, by using collection keybindings, or by pinning them. Additionally, users can control which concepts are pinned here. Concepts in the collection can be dragged to other components, such as the highlight panel, to create a matching rule; the primary visualization, to refocus; or the query builder, to add them to a query. Collections save users from having to remember full concept names and allow rapid return to points of interest—for instance, quickly relocating the view to \nit{Abnormality of the head}.

The details panel (Figure~\ref{fig:concept-details}) displays information for the currently selected concept. Depending on the ontology, the number of attributes varies. Most ontologies' concepts include labels, IRIs, parents, and children. In HPO, for example, concept definitions add clarity to a term’s meaning. While the primary visualization provides high-level spatial information, it forgoes details in favor of visual clarity. As such, the details panel delivers essential low-level information for individual concepts.

\subsubsection{Query Building Panels}
The second section of the right region contains the query building panels (Figure~\ref{fig:overview}.E), which allow users to create, resolve, and analyze queries. Navigating an ontology hierarchically only reveals part of its structure and semantics, and inspecting concepts one at a time can be inefficient. Query building enables non-linear exploration, helping users verify inferences or discover related concepts. We discuss this in-depth in our usage scenario (Section~\ref{sec:floqe-sce}).

\begin{figure}[htb]
    \centering
    \begin{subfigure}[t]{.30\textwidth}
        \centering
        \captionsetup{width=.9\linewidth}
        \setlength{\fboxsep}{0pt}\fbox{\includegraphics[width=0.9\linewidth]{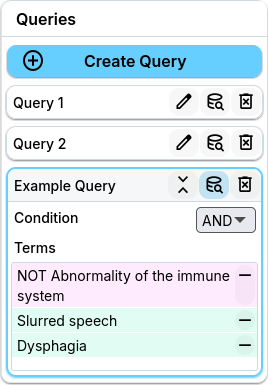}}
        \caption{The query builder panel where users can construct queries from concepts or other queries.}
        \label{fig:query-build}
    \end{subfigure}
    \begin{subfigure}[t]{.30\textwidth}
        \centering
        \captionsetup{width=.9\linewidth}
        \setlength{\fboxsep}{0pt}\fbox{\includegraphics[width=0.9\linewidth]{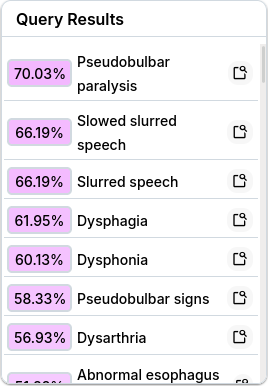}}
        \caption{The query results panel showing the similarity between the created query and ontology concepts.}
        \label{fig:query-result}
    \end{subfigure}
    \caption{The query building panel of the rightmost region of \floqe. Used for creating queries and viewing the results.}
    \label{fig:right-overview-1}
\end{figure}

Queries are created in the query builder control (Figure~\ref{fig:query-build}) from concepts, other queries, and standard logical operators (AND, OR, NOT), corresponding to intersection, union, and negation. Once finalized, the front-end sends the query to the back-end resolver, which converts it to a composite concept. This composite concept’s embedding is compared with primitive concept embeddings, returning the top-\(k\) most similar concepts. Queries may contain any concepts, including logically inconsistent combinations---which, while ineffective when resolved with traditional embeddings, can still yield meaningful results with our fuzzy ontology embeddings.

Results are displayed in the query results panel (Figure~\ref{fig:query-result}), showing similarity scores for each primitive concept. Users can expand results for details, or drag them into new queries or other components. The similarity stain option in the visualization control panel colors the primary visualization according to these similarity values. The query resolver uses cosine similarity, and because primitive concept embeddings are stored in a vector database, queries resolve nearly instantly.

\subsection{Usage Scenario} \label{sec:floqe-sce}
The following scenario illustrates how \floqe\ supports ontology exploration and the practical applications of fuzzy logic queries. We use the HPO for this scenario. We assume the user knows basic medical terms but is unfamiliar with HPO. The embedding used is an \(\alpha\)-embedding (as described in Section~\ref{sec:hierarchical-interpretation}) with \(\alpha = 0.25\). For fuzzy operators, we use the product t-norm for conjunction, its t-conorm for disjunction, and the standard negator for negation.

A physician, henceforth \textit{the user}, has a patient who has difficulty speaking and swallowing, with no history of immune disorders, and no signs of infection. To help this patient, the user has two goals. One, they want to identify possible conditions that could explain the symptoms. Two, they wish to learn about any secondary ailments that the patient may experience now or in the future. The user knows that HPO contains information about abnormal conditions, but its size and complexity necessitates the use of an exploratory tool. Thus, the user launches \floqe\ to explore HPO and help their patient.

After \floqe\ loads HPO, its front-end---displaying a nested-treemap as the primary visualization---presents the ontology to the user. The user starts their exploration by adjusting the settings in the visualization control panel. First, the user sets cell scaling to be equal, as the relative size of concepts is irrelevant to the their goals. Second, the user sets the tiling so that labels are clearly visible. Lastly, they set visible depth to a low value---HPO has many terms and seeing too many layers is overwhelming. Afterwards, the user creates highlight rules to stain concepts containing \textit{voice}, \textit{speech}, or \textit{swallowing} for quick identification. With the user's initial setup complete, they can now quickly identify concepts of interest. Users of \floqe\ can, with little effort, set up the primary visualization in ways that streamline their future searches.

Seeing the concept \nit{Phenotypic abnormality}, the user selects it and begins their search. During their search, the user encounters concepts related to their patient's condition. These concepts are pinned (for easier identification) and added to the concept collection for use in queries. To explore more specific concepts, the user activates focus mode to drill into \nit{Phenotypic abnormality}’s children. They see \nit{Abnormality of the voice}, and although it does not contain what they are looking for, it is collected for query use regardless. Additionally, the user spots \nit{Abnormality of the immune system} and adds it to the collection, to be negated in future queries.

Wanting to quickly find something, the user inputs \textit{speech} into the search bar. Among the results is the concept \nit{Slurred speech}, which matches the patient's symptoms. Selecting it takes the user deeper into the ontology, where they then pin it. Unsure of where they are, the user toggles on the depth numbers and sees that \nit{Slurred speech} is eight layers deep and has no children. Then, the user increases the depth to view the full ancestry and spots that Slurred speech is within the \nit{Abnormality of the nervous system} subtree. The user selects this broader concept, reduces depth for clarity, and continues exploring. 

Already the user has started to identify landmark concepts related to their problem, has learned the routes between them, and is now aware of the neighborhood in the ontology that contains concepts of interest. By quickly identifying this information, the user efficiently narrows the search space. \floqe's support for rapid, seamless navigation allows users to quickly see concept contents at a glance and freely navigate large ontologies.

The user’s exploration continues until, within \nit{Abnormality of the nervous system} child concept \nit{Abnormal nervous system physiology}, they encounter \nit{Dysphagia}. The focus mode alternative views and concept details panel show the user that \nit{Dysphagia} is the medical term for difficulty swallowing. Combined with the earlier discovery of \nit{Slurred speech}, the user feels ready to construct queries. The user formulates their query as the presence of Slurred speech and Dysphagia while excluding \nit{Abnormality of the immune system}. They do this by dragging concepts from the collection to the query building panel to form the composite concept \(Q_1\):%
\begin{equation*}
Q_1 \equiv \nit{Slurred speech} \sqcap \nit{Dysphagia} \sqcap \neg \nit{Abnormality of the immune system}.
\end{equation*}

\begin{figure}[htb]
    \centering
    \setlength{\fboxsep}{0pt}\fbox{\includegraphics[width=\linewidth]{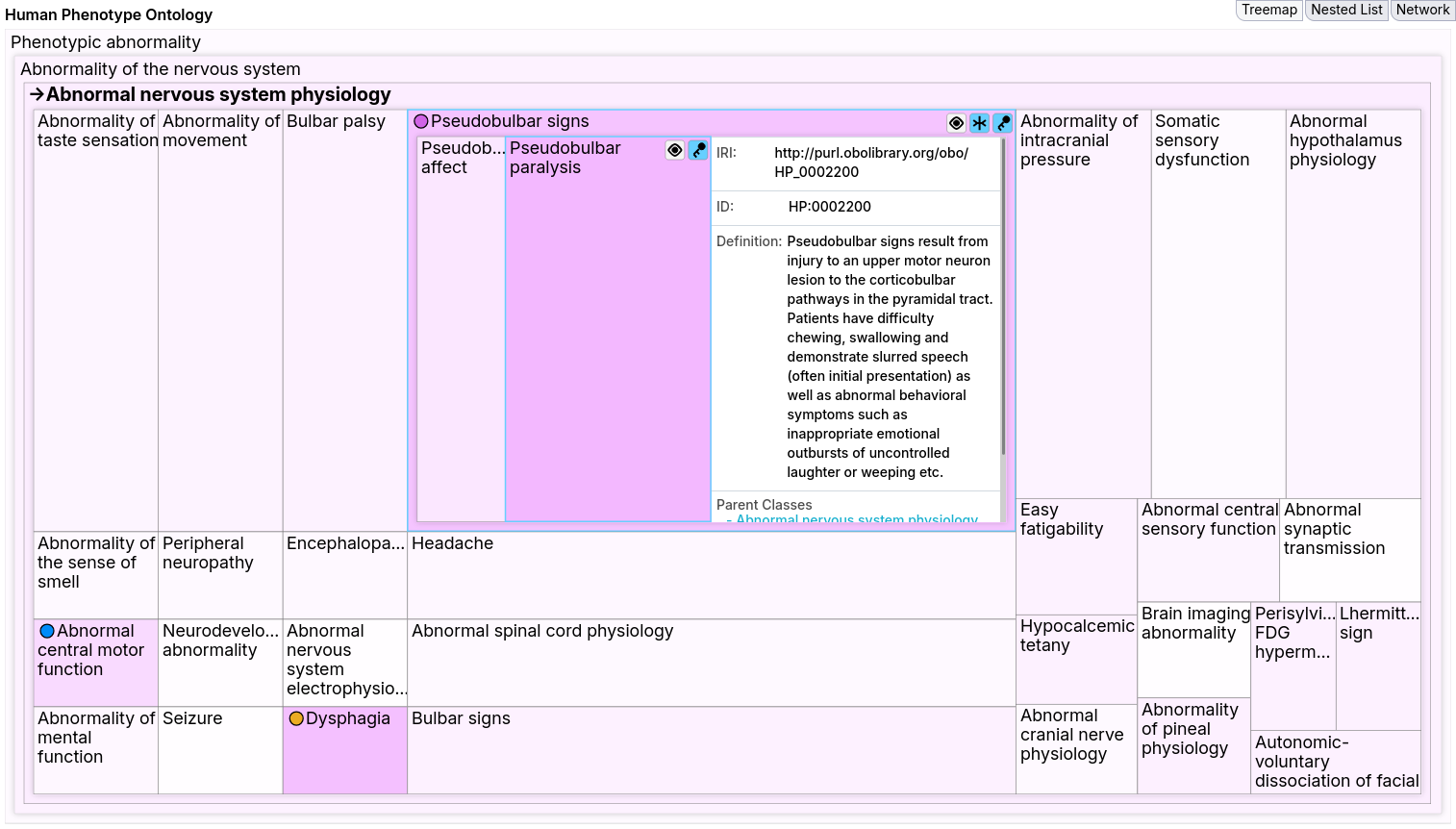}}
    \caption{The primary visualization set to the treemap view. The similarity stain has been turned on and the concept \nit{Pseudobulbar signs} has been focused, locked, and split to show details. Its child \nit{Pseudobulbar paralysis} is also focused and locked.}
    \label{fig:scenario}
\end{figure}

The user submits their query, \floqe's back-end instantly resolves it, and the front-end updates to show the results. Among these results are concepts such as \nit{Pseudobulbar paralysis}, \nit{Pseudobulbar signs}, and \nit{Abnormal esophagus physiology}. Wanting to locate these concepts, the user enables the similarity stain and spots the highlighted \nit{Pseudobulbar signs}. Using the focus mode, the user locks and splits \nit{Pseudobulbar signs} to expand it and see its details. Notably, this reveals \nit{Pseudobulbar paralysis} within \nit{Pseudobulbar signs}. Figure~\ref{fig:scenario} shows the combined effect of this similarity stain and focus mode, highlighting the pseudobulbar concepts. These pseudobulbar concepts point to neurological causes that explain the patient’s symptoms. On the other hand, \nit{Abnormal esophagus physiology} offers an alternative explanation for the symptoms.

These results have now narrowed down potential causes that the user can investigate to diagnose their patient. Such concepts would have been difficult to find with keyword search, as the user's vocabulary does not match the ontology’s precisely. Moreover, the user would have been unable to find these concepts with SPARQL queries for this same reason. With \floqe, users build queries visually by dragging and dropping concepts and other queries, without needing to type exact labels. Furthermore, nesting queries helps users separate the components of their queries into understandable segments and construct complex concepts. By supporting fuzzy queries, \floqe\ enables users to find relevant concepts even when their requests are imprecise.

An example of an imprecise query is the user’s search for secondary ailments. The user creates this query by updating their original one to exclude \nit{Abnormality of the voice}. In a non-fuzzy system, excluding \nit{Abnormality of the voice} while searching for conditions related to speech and swallowing might seem counterintuitive or overly restrictive. However, because the \(\alpha\)-embedding captures fuzzy membership across the ontology, the query can still yield useful insights. The new query narrows the focus to concepts returned by the original that are unrelated to \nit{Abnormality of the voice}. The updated query \(Q_2\) is:
\begin{multline*}
Q_2 \equiv \nit{Slurred speech} \sqcap \nit{Dysphagia} \sqcap \\
\neg (\nit{Abnormality of the immune system}  \sqcup \nit{Abnormality of the voice}).
\end{multline*}

Submitting \(Q_2\) returns concepts related to digestive issues, which is expected since \nit{Dysphagia} belongs to the \nit{Abnormal esophagus physiology} subtree under \nit{Abnormality of the digestive system}. This suggests the patient’s condition may affect the digestive system. The fuzzy ontology embeddings preserve such relationships even in complex composite concepts. Armed with this knowledge, the user can further question their patient to narrow down their condition. The user asks about irritability and emotional outbursts to check for \nit{Pseudobulbar signs}, and inspecting the throat to check for \nit{Abnormal esophagus physiology}. 

This scenario demonstrates how \floqe’s features and fuzzy ontology embeddings enable effective ontology exploration. Without prior knowledge of HPO’s structure or query languages like SPARQL, the user was able to find relevant concepts and understand their relationships. The interface’s interactive visualizations allows the user to build queries visually, while fuzzy logic supports the creation of useful queries even without expert knowledge. Together, these capabilities present a flexible and user-friendly approach to exploring ontological information.
\section{Conclusion and Future Work}\label{sec:conclusion}

This paper presented \floqe, a prototype system for visual query building over ontologies using fuzzy logic and membership-based embeddings. \floqe\ enables users to define complex, potentially vague concepts through familiar set operations—intersection, union, and negation—and interprets these as fuzzy sets embedded in a continuous vector space. This representation supports efficient, similarity-based query evaluation. Through an interactive visual interface, \floqe\ integrates search, navigation, query construction, and result presentation, making query formulation more accessible and intuitive, particularly for non-expert users. By combining symbolic and sub-symbolic techniques, the system supports approximate reasoning and flexible concept retrieval, in contrast to traditional methods that rely solely on formal syntax and strict logic.

As a prototype, \floqe\ offers several avenues for future development. First, improving performance and scalability for very large ontologies remains essential, requiring optimization of both front-end visualizations and back-end embedding computations. Second, expanding functionality—such as adding collapsible tree views or new panels for organizing and comparing query results—could better support diverse user needs. Third, investigating alternative embedding techniques, especially those optimized for specific ontology types or designed for greater interpretability, may yield improved results. Finally, extending \floqe\ to handle more expressive ontologies with richer axioms beyond hierarchical subsumption would require enhancements to both the query builder and embedding mechanism, enabling support for a broader range of logical constructs and relationships.

\bibliographystyle{plain}
\bibliography{bibliography}

\end{document}